\documentclass[runningheads]{llncs}
\usepackage[T1]{fontenc}
\usepackage{xurl,hyperref}
\usepackage{amsmath,amssymb}
\usepackage[table]{xcolor}
\usepackage{multirow}
\usepackage{comment}
\usepackage{listings}
\usepackage{graphicx}
\usepackage{color}

\begin{document}
\title{Technical Evaluation of a Disruptive\\ Approach in Homomorphic AI\thanks{This is the extended version of the talk presented at CyberWiseCon 2025 in Vilnius, Lituania in May 21$^{st}$-23$^{rd}$, 2025.}}

\author{Eric FILIOL} 
\authorrunning{E. Filiol}
\institute{Retired Professor - Independent Researcher\\
\email{efll@protonmail.com}\\
\url{https://ericfiliol.site}}
\maketitle             
\begin{abstract}
We present a technical evaluation of a new, disruptive cryptographic approach to data security, known as HbHAI (Hash-based Homomorphic Artificial Intelligence).
HbHAI is based on a novel class of key-dependent hash functions that naturally preserve most similarity properties, most AI algorithms rely on.  
As a main claim, HbHAI makes now possible to analyse and process data in its cryptographically secure form while using existing native AI algorithms without modification, with unprecedented performances compared to existing homomorphic encryption schemes.

We tested various HbHAI-protected datasets (non public preview) using traditional unsupervised and supervised learning techniques (clustering, classification, deep neural networks) with classical unmodified AI algorithms. This paper presents technical results from an independent analysis conducted with those different, off-the-shelf AI algorithms. The aim was to assess the security, operability and performance claims regarding HbHAI techniques. As a results, our results confirm most these claims, with only a few minor reservations.
\end{abstract}
\section{Introduction}
In this paper, the “AI” term is used to describe all data analysis techniques (machine learning, deep learning, big data) to the exclusion of LLM (generative AI). For sake of concision, Homomorphic AI must be understood as Homomorphic Encryption for AI. 

As far as AI is concerned, most approaches require to use third-party environments such as clouds which provide suitable tools. The only other possibility is to use ``on-premises'' environments with suitable and skilled enough tech teams to deliver, run, maintain and to operate them. With cloud solutions the main security drawback lies on the fact that data owners do no longer control the access to data. 
Outsourcing data for the purposes to use dedicated AI tools as a service thus represents either a weakness or a danger. Indeed, we observe that a data has three essential ``vocations'':
\begin{itemize}
	\item to grow indefinitely (cost issues in terms of storage, computing time, bandwidth consumption),
	\item to be shared or accessed (and thus lead to misuse) with dubious third-parties (e.g. data brokers, national police or intelligence agencies\footnote{The major players in the cloud \url{https://dgtlinfra.com/top-cloud-service-providers/} are ALL subject to restrictive national laws requiring them to allow their intelligence services to access data passing through/stored in their environments in various ways, without being authorized to alert the owners of the data concerned.}),
	\item and, worse, to leak in the wake of attacks (for 2024, for instance refer to~\cite{philmun}). It is worth noticing that this risk equally exists for ``on-premises'' environments.
\end{itemize}
The most effective protection is to be able to process data directly in encrypted form without using data under their plaintext form. In this way, in storage or during processing, any attacker or unauthorised third party will only have access to data in a form that cannot be used by them. This protection is called \textit{Homomorphic Artificial Intelligence} (HAI) coming from the original research area of \textit{Homomorphic Encryption} (HE). 

At the end of 2020, we launched a collaborative project with an Estonian start-up, Hope4Sec, to develop a totally new and disruptive approach to homomorphic encryption applied to AI. The aim was to start from scratch and design a homomorphic data analysis scheme, called HbHAI (standing for \textit{Hash-based Homomorphic Artificial Intelligence}) that would provide at least the same level of cryptographic security for the data, while removing the constraints and limitations previously mentioned for existing HE schemes. In particular, the aim was both to significantly reduce the ecological footprint and to enable existing AI algorithms to be used without having to modify or rewrite them. 

The first primitives were finalised at the end of 2023 and the formalisation work, based on specifications, was published in 2025~\cite{filiol0}. Hope4Sec has produced several datasets for testing and evaluation, which have been made available to us as non public previews. These datasets will be published during 2025~\cite{sepp}.

For the time being, the HbHAI scheme is not public because it is not yet protected in terms of intellectual property. Moreover the industrial exploitation is still pending. 

In this paper, we detail the results presented at CyberWiseCon 2025~\cite{filiol1} on the two datasets provided by Hope4Sec for evaluation. 
The aim of our analysis was
\begin{itemize}
	\item to evaluate Hash-based Homomorphic Artificial Intelligence (HbHAI) techniques on two encrypted datasets provided,
	\item to perform an initial assessment of cryptographic security claims with a black-box analysis (we do not have access of the mathematical description of HbHAI to date),
	\item to check the claims about dataset size and computing time reduction,
	\item and finally to check the operability claims (off-the-shelf tools in their native form, refer to Subsection~\ref{subsec32}. 
\end{itemize}
The paper is organised as follows. In Section~\ref{sec1}, we present the current landscape of HE techniques to date. In Section~\ref{sec2}, we explore the current classic HE techniques used specifically for AI and we shortly present HbHAI techniques and their main general features and specifications. In Section~\ref{sec3}, we present the analysis of HbHAI datasets and evaluate the claims with respect to HbHAI features. Finally Section~\ref{sec4} summarizes our results and mentions the future directions that HbHAI should know and follow.

\textbf{Disclaimer.} \textit{Any views, opinions and materials presented in this paper are personal and are the results of my own research work.
They belong solely to the speaker and, in any case, they do not represent those of people, institutions, companies or organizations that 
the speaker may or may not be associated with in professional or personal capacity (including past, present and future employers).}
%%%%%%%%%%%%%%%%%%%%%%%%%%%%%%%%%%%%%%%%%%%%%%%%%%%%%%%%%%%%%%%%
\section{Homomorphic Encrypption Landscape}\label{sec1}
%----------------------------------------------------------------
\subsection{State-of-the-Art of HE}
A lot of systems have been produced but all of them can be seen in a way or another an extension of public key cryptography~\cite{ieee1}. 
All known systems use a combination of private and public keys and are built on primitives used in public key cryptography. Homomorphic 
cryptography is based on the concept of homomorphism in mathematics~\cite[pp. 189ff]{kostrikin}, \cite{lobo2020}. Homomorphisms are maps 
between algebraic structures that preserve a number of operations, thus maintaining the same overall structure. The most optimal category 
of HE techniques are \textit{Fully Homomorphic Encryption} (FHE). 

This area of research emerged in 1978 and has become a critical area of research with the explosion and widespread use of the cloud. The 
spread and development of AI on third-party cloud platforms has made this research even more vital.

Three main types of HE does exit nowadays. They are determined by the ability to break down and express a problem (formalised by the 
circuit concept) into a sequence of atomic operations (addition, multiplication, permutation\ldots) formalized of gates in circuits.
\begin{itemize}
	\item \textbf{Partially homomorphic encryption (PHE)} supports the evaluation of circuits consisting of only one type of gates 
	(e.g. addition or multiplication).
	\item \textbf{Somewhat homomorphic encryption (SHE)} can evaluate two types of gates, but only for a subset of circuits (not all problems)
	\item \textbf{Fully homomorphic encryption (FHE)} allows the evaluation of arbitrary circuits composed of multiple types of gates of unbounded depth (strongest notion of HE).
\end{itemize}
%-----------------------------------------------------------------
\subsection{Drawbacks and limitations}
A certain emergency linked to the development of the industry (cloud, AI) has encouraged research to consider already existing cryptographic solutions, which has hampered the search for innovative solutions. The powerful and elegant public key cryptographic solutions used for certain security functions (authentication, signature, encryption of small amounts of information, optimised key management\ldots) are proving to be unsuitable for naturally intensive use in the context of AI.

Because of their limitations, known HE techniques are still of limited operational use, particularly for AI (see \cite{jkun,lin} for a synthetic summary):
\begin{itemize}
	\item Security:  HE schemes are inherently \textit{malleable}. It means that it is possible to transform a given ciphertext $C$ into another ciphertext $C'$ which decrypts to a related plaintext \cite{dolev}. That is, given an encryption of a plaintext $M$, malleability makes possible to generate $C'$ which decrypts to an admissible plaintext $f(M)$ for a known function $f$ without necessarily knowing or learning $M$. In terms of malleability, HE schemes have weaker security properties than non-homomorphic schemes.
	
	\item Performance: HE is computationally intensive, especially for large datasets or complex computations. This results in significant processing overhead, in resource intensive and performance slowdown. For instance, running FHE on CPU is at least a million times slower than the corresponding unencrypted program \cite{jkun}.
	\item Complexity: HE requires mastering advanced mathematical concepts and algorithms as well a secure programming. This can make it challenging for developers and engineers to implement, make evolve, adapt and scale HE systems. All AI native algorithms have to be totally rewritten to produce a HE-compliant version. The implementation of cryptography is a critical skill to be sure that cryptographic security is not weakened. Additionally, HE requires strong key management processes to ensure the security of the encrypted data.
	\item The cryptographic security requirements imply rather strong limitations on what sorts of optimizations you could use because realizing that optimization would necessarily break the cryptography \cite{jkun}. 
	\item Bandwidth and storage concerns. HE schemes generally significantly increase the size of the data being encrypted. This has a significant impact on computing performance and therefore on energy consumption.
	\item Key Management. Asymmetric keys used in HE schemes have a large size (a few Mb in average) which makes them complex to manage. In a industrial, real-life context, the use of Hardware Security Module (HSM) is still difficult not to say unpracticable. 
\end{itemize}

Energy demand and the ecological impact of AI are a growing concern. While the alarmist forecasts for 2024 seem to need to be moderated according to the IEA \cite{iea}, the widespread use of current HE techniques will contribute to a significant increase in this energy and ecological footprint.
%%%%%%%%%%%%%%%%%%%%%%%%%%%%%%%%%%%%%%%%%%%%%%%%%%%%%%%%%%%%%%%%
\section{Homomorphic Encryption for Artificial Intelligence} \label{sec2}
Fully homomorphic encryption (FHE)~\cite{fhe} is the most accomplished homomorphic encryption technique (for a detailed presentation of HE systems, the reader can refer to~\cite[Chap. 2]{he4ds}). FHE provides a powerful scheme for privacy-preserving computation of externalized data. 

By enabling atomic operations such as permutations, additions and multiplication\ldots directly on ciphertext, and provided that the underlying problem can be decomposed into a sequence of such operations, FHE enables secure data processing without the need for decryption. This capability makes FHE particularly suitable for a wide range of database operations, including statistical analysis and aggregation queries~\cite{hades}. However, FHE alone cannot address the full range of functionalities, as it inherently lacks the ability to compare ciphertexts (a critical requirement for many database operations). 

The few existing homomorphic system for Data Science use FHE systems as a black-box and builds layers of knowledge on top of it. The existence of an efficient FHE implementation (software or hardware) is assumed and the key study is generally how it can be utilized to build different AI and data science applications. 
%--------------------------------------------------------------
\subsection{A Short Survey of Existing HE Schemes (CKKS \& BFV)}
FHE schemes are based on the seminal work of Craig Gentry~\cite{fhe} and allow users to evaluate any polynomial function or binary circuit over encrypted data.

Two widely most used homomorphic encryption schemes are the Brakerski/Fan-Vercauteren (BFV) scheme~\cite{bfv} and the Cheon-KimKim-Song (CKKS) scheme~\cite{ccks}. Both schemes leverage the hardness of the Ring/Torus Learning with Errors (R/TLWE) problem~\cite{rlwe} to ensure cryptographic security.
\begin{description}
\item[BFV Scheme] The BFV scheme is designed for exact arithmetic over encrypted integers. It is particularly well-suited for applications where the precision of computations must be preserved (encrypted database queries, voting systems and secure machine learning). BFV operates efficiently by supporting addition and multiplication over ciphertexts while maintaining accurate results.

\item[CKKS Scheme] This scheme is tailored for approximate arithmetic over real numbers, making it ideal for applications such as encrypted signal processing, privacy-preserving AI and financial computations. CKKS introduces a trade-off between precision and computational efficiency, allowing for flexible scaling in real-world scenarios.
\end{description}
A third scheme, called TFHE (standing for Fully Homomorphic Encryption over the Torus) has been introduced in~\cite{tfhe}. This can be seen as a particular case of FHE where the bootstrapping has been very significantly optimized. The torus is the underlying additive group of
LWE denoted $\mathbb{T}$ and defined as $\mathbb{T} = \mathbb{R}/\mathbb{Z}$.

One of the properties of these two FHE schemes is that they add noise to a ciphertext during the encryption process. Homomorphic operations (especially multiplication) increase the noise. If the noise becomes too great, the ciphertext may be indecipherable. BFV et CKKS schemes enables bootstrapping which is the procedure of ``refreshing'' a ciphertext by running the decryption function homomorphically on it, resulting in a reduced noise. 

However bootstrapping brings itself a number of issues. It involves additional bootstrapping keys (of rather large size, for instance 16Mb for TFHE~\cite{tfhe}). A more or less important overhead is introduced. But the main issue may be a security concern since the process, being itself homomorphic, requires a local computation with additional public key~\cite{fhe}. In a external, multi-tenant environment, the risk of side channel attacks or fault injection attacks cannot be excluded (see for instance~\cite{fheinjatt,zama}). 

Consequently an ideal homomorphic encryption scheme requires that absolutely no cryptographic operation of any kind (such as bootstrapping) involving cryptographic keys is performed at any time, once the initial plaintext encryption has occurred (in the user's secure environment).

Optimized implementations of BFV and CKKS schemes can be found in HElib~\cite{helib}, TenSEAL~\cite{tenseal}, SEAL~\cite{seal} and OpenFHE~\cite{openfhe} librairies. An efficient implementation of TFHE is also available in~\cite{tfhelib}. Performances comparisons can be found in~\cite{comp4,comp2,comp5,comp1,comptfhe,comp3}.
%----------------------------------------------------------------
\subsection{Homomorphic AI Hash Functions (HbHAI)} \label{subsec32}
AI techniques cannot consider classical cryptographic hash functions to provide homomorphic capabilities. This is the reason why a new class of keyed-hash functions has been imagined. The use of keyed hash functions aims at the same to provide a strong cryptographic security and a significant data size and computing time reduction for the AI algorithms.
 
 \begin{definition} (HAI Hash Function Class) \cite{filiol0} \label{defhai}
A keyed hash function for HAI applications is a function $H_{K, \delta}$ parametrized by a secret key $K$ and a compression rate $\delta$, which has, as a minimum, the following two properties:
	\begin{enumerate}
		\item \textbf{Compression} — $H_{K, \delta}$ maps an input $x$ of arbitrary finite bit length $n$, to an output $H_{K, \delta}(x)$ of bit length $\frac{n}{\delta}$.
		\item \textbf{Ease of Computation} — Given $H_{K, \delta}$ and an input $x$, $H_{K, \delta}(x)$ is easy to compute.
		\item \textbf{Similarity Preserving} - For a given similarity measure $S$ and any three objects $x, x', x''$ then we have, 
		\[S(x, x'') < S(x, x') \Leftrightarrow S(H_{K, \delta}(x), H_{K, \delta}(x'')) < S(H_{K, \delta}(x), H_{K, \delta}(x'))\]. 
	\end{enumerate}
\end{definition}
This definition considers similarity instead of the more restricting concept of distance. Most AI techniques, not to say all, are based in a way or another on the central concept of similarity (between objects). Most similarity measure can be converted to distance but not all (for instance Cosine similarity). 

Let us consider which threat model (with respect to AI) we have to consider and which security properties must be fulfilled. 
\begin{definition} (HAI Hash Function Security)~\cite{filiol0} \label{hai_sec}
For a HAI hash function $H_{K, \delta}$ with inputs $x, x'$ and outputs $y, y'$, the required cryptographic security properties are:
\begin{enumerate}
\item \textbf{Sample Space Security (SSS)} - It must be computationally impossible to recalculate inputs and to determine their configuration and nature. This implies pre-image resistance of classic cryptographic hash functions.
\item \textbf{Model Security (MS)} - It is not possible to extract the model or produce an equivalent model. In other words, it is impossible to produce an identifiable model from the different values $H_{K, \delta}(x_i)$. This implies the collision-resistance and the second-preimage resistance of classic cryptographic hash functions.
\item \textbf{Non Malleability} - It must not be possible to modify $H_{K, \delta}(x)$ in order to produce an admissible value $H_{K, \delta}(x')$ with respect to the underlying model.
\end{enumerate}
\end{definition}
The use of hash functions (non-injective transformations) invalidates the concept of decryption in HE. With HbHAI, a different approach has been defined. Only the owner of the data knows the correspondence between the unencrypted and encrypted versions of an object/individual in a dataset, thanks to the index of each individual (after decrypting the index eventually).
It is possible \cite{filiol0} to transpose the results of AI algorithms obtained on the encrypted version of a dataset to the unencrypted version of the dataset, using only the indexes (clustering, classification\ldots). The decryption operation is therefore no longer necessary, in the specific context of AI.

In addition to strong security requirements, HbHAI specifications also include other requirements, the most essential of which are as follows: 
\begin{description}
	\item[Frugality] \textit{i.e.} minimizing the computational resources required both to calculate the models and to operate them. This concerns both a significant reduction in data size (parameter $\delta$) and a significant reduction in computation time. The main benefit is reduced energy and ecological footprints and an enhanced operability in critical and constrained systems (drone, embedded systems, IoT).
	\item[Portability] Data protected by HbHAI must to be processed by existing algorithms, in their original form, without rewriting (Keras, Tensorflow, custom implementation of classic AI algorithms\ldots).
	\item[Sovereignty and Independence] HbHAI technology must be operated in constrained environments and not on cutting-edge technologies (subject to US embargoes and export controls).
\end{description}   
%%%%%%%%%%%%%%%%%%%%%%%%%%%%%%%%%%%%%%%%%%%%%%%%%%%%%%%%%%%%%%%%%%%%%%%%%%%
\section{AI Analysis of HbHAI-protected Datasets} \label{sec3}
Two datasets have been provided with the following parameters: a 256-bit key $K$ and $\delta \in  [3, 6] \subset \mathbb{R}$. These two datasets are presented in \cite{sepp} and should be published in July on Hope4Sec's website and Kaggle\footnote{\url{http://kaggle.com/}}. They have been specifically design to evaluate most AI algorithms and problems.   

Table~\ref{tab1} summarizes the main features of those two datasets~\cite{sepp}.
\begin{table}
	\centering
	\begin{tabular}{|c|c|c|} \hline \rowcolor{lightgray} 
		Features                                           &       Dataset 1          &           Dataset 2           \\ \hline \hline
		Data type                                          &  \; \;Cyber data \; \;   &   \; \;Greyscale Images\; \;  \\ \hline
		Number of objects (training)                       &        2,000             &            60,000             \\ \hline
		Number of objects (validation)                     &          200             &            10,000             \\ \hline
		Number of features                                 &        49,955            &              N/A              \\ \hline
		Number of clusters/classes                         &           2              &              10               \\ \hline
		\; \;Original dataset size (Tr.+Val.)\; \;         &         14 Mb            &           30.3 Mb             \\ \hline
		\; \;HbHAI-protected size (Tr.+Val.)\; \;          & \multirow{2}*{4.70 Mb}   &      \multirow{2}*{11.3 Mb}   \\ 
		            ($\delta = 3$)                         &                          &                               \\ \hline
		\; \;HbHAI-protected dataset size (Tr.+Val.)\; \;  & \multirow{2}*{N/A}       &     \multirow{2}*{5.2}        \\  
		            ($\delta = 6$)                         &                          &                               \\ \hline
	\end{tabular}
	\medskip
	\caption{Features of Datasets Provided}	\label{tab1}
\end{table}
All of analysis and testing have been performed on constrained environment to evaluate HbHAI specification regarding operability:
	\begin{itemize}
	\item Odroid H4 Ultra single board computer with 8-core CPU (architecture Alder Lake N) 
	\item Linux Pop!\_OS with 32 Gb DDR5-4800 + 1Tb SSD.
	\item Linux OpenSuse 15.4  with 32 Gb DDR5-4800 + 1Tb SSD (as a development environment with gcc + GMP).
\end{itemize}
The operational conditions of our analysis 
	\begin{itemize}
	\item We had access only to the unprotected version (plaintext) and HbHAI-protected version of datasets.
	\item We did not have any knowledge on the HbHAI primitives/techniques themselves (blind analysis).
	\item We want to assess a comparative data analysis with respect to performances (accuracy, computing time).
	\item We want to conduct a black-box analysis of cryptographic security with respect to the threat model defined in Definition~\ref{hai_sec}.
\end{itemize}
%%%%%%%%%%%%%%%%%%%%%%%%%%%%%%%%%%%%%%%%%%%%%%%%%%%%%
\subsection{Use-Case 1: Clustering/Classification of Cybersecurity Data}
This dataset gathers non public data coming from the cybersecurity domain. Only the HbHAI-protected version of the dataset was available to us.
Each individual (object) $u_i \in \mathcal{U}$ is described by 49,955 different characteristics or features $j$ (categorical features).
 
This two-class dataset is intended to test unsupervised learning (clustering) but also classification (identifying to which class new objects belong). 
The dataset is made up of
\begin{itemize}
\item One training set (objects) containing 2,000 files denoted \texttt{training-xxx-k} where \text{xxx} denotes the individual ID and \text{k} describes its class ($k \in \{1, 2\}$). Each set contains 2,620 bytes.
\item One validation set (objects) containing 200 files denoted \texttt{validation-xxx-k} where \text{xxx} denotes the individual ID and \text{k} describes its class ($k \in \{1, 2\}$). Each set contains 2,620 bytes.
\end{itemize}

The use of existing software such as scikit.klearn or R are not very practical when processing individuals with many features (significant overhead). Moreover they do not allow (at least easily) a precise evaluation of code execution and speed-up. So we decide to implement high-speed classical k-modes algorithm~\cite{huang1998} for the clustering and $k$-NN algorithm~\cite{fix1951} for the classification (supervised part), from scratch in C language as follows:
\begin{itemize}
\item Use of GNU Multiple Precision Arith. Lib. (GMP~\cite{gmp}). Aside the powerful environment for multi precision, GMP includes accurate 
      and easy-to-use profiling instructions natively.
\item Each individual is described by a unique 49,955-bit (plaintext) or 17,000-bit (HbHAI protected) integer.
\end{itemize}
The following C code extract show how simple it is to load and process individuals with a large number of features. GMP implementation enables to work only with bitwise operations on very large integers. Note that this code is agnostic of the nature of data (encrypted or not)
\begin{lstlisting}[language=C]
/* Variable allocations */
mpz_init(vect); mpz_init(ccenter1); 
mpz_init(ccenter2);
for(i = 0; i < nbvect;i++) mpz_init(vectset[i]);
[...]
	
/* Import data to be clustered */
for(i = 0; i < nbvect;i++) 
mpz_import(vectset[i], countp, 1, 1, 1, 0, data);
	
/* Compare an individual with current cluster center1 */
mpz_and(vect, vectset[j], ccenter1):
cdist1 = (float)(mpz_popcount(vect));
\end{lstlisting}

This very classic clustering program was run on plaintext data and on HbHAI-protected data. We thus obtained two partition, $P = (P_1, P_2)$ on plaintext dataset and $P' = (H_1, H_2)$ on HbHAI-protected data. It was then necessary to compare the two partitions to evaluate the impact of HbHAI on clustering.

For that purpose, we use the classical Rand Index~\cite{rand1971}. The Rand index has a value in $[0, 1] \subset \mathbb{R}$ with 0 indicating that the two data clusterings do not agree on any pair of points and 1 indicating that the data clusterings are exactly the same.

With an equal number of iterations, we systematically obtained $R = 1$, which means that HbHAI perfectly preserves data clustering.
Regarding classification, the $k$-NN algorithm (supervised classification) gives exactly the same results with and without HbHAI.

As a general result we confirm the following:
\begin{itemize}
\item A simple data size comparison (see Table~\ref{tab1}) the effective datasize reduction is equal to $2.99$ (observed). 
\item The computing time speed-up is equal to $2.8$.
\item Compared to the theoretical value $\delta = 3$, the experimental results obtained on this dataset are in line with theory announced. 
\end{itemize}
%%%%%%%%%%%%%%%%%%%%%%%%%%%%%%%%%%%%%%%%%%%%%%%%%%%%%%%
\subsection{Use-case 2: Zalando's Fashion-MNIST Reference Dataset}
Fashion-MNIST~\cite{dataset2} is a dataset of Zalando's article images—consisting of a training set of 60,000 examples and a test 
set of 10,000 examples. Each example is a 28x28 greyscale image, associated with a label from 10 classes. Fashion-MNIST 
dataset's purposes is to serve as a direct drop-in replacement for the original MNIST dataset \cite{mnist} for benchmarking 
machine learning algorithms. It shares the same image size and structure of training and testing splits. 

From this public dataset, two versions protected with HbHAI were available for the test:
\begin{itemize}
\item One version with $\delta = 3$. The set contains the same files as in the original dataset but each 28x28 greyscale image is compressed as a 16x16-byte array. 
\item One version with $\delta = 6$. The set contains the same files as in the original dataset but each 28x28 greyscale image is compressed as a 132-byte row.
\end{itemize} 
For each of the ten classes $j = \{1, 2, \ldots, 10\}$, the encryption key is also used to perform a permutation $\pi^j_K$ of the individuals before encrypting them. Individual $i$ in the class $j$ thus becomes individual $\pi^j_K(i)$ in class $j$.

This dataset is intended to test most recent AI algorithms and especially Neural Networks~\cite{dataset2}.
%---------------------------------------------------------------------------------------------------------
\subsubsection{Test with TensorFlow/Keras ($\delta = 3$) - MLP Approach}
\begin{figure}[h]\label{fig1}
	\begin{center} 
		\includegraphics[width=12.3cm]{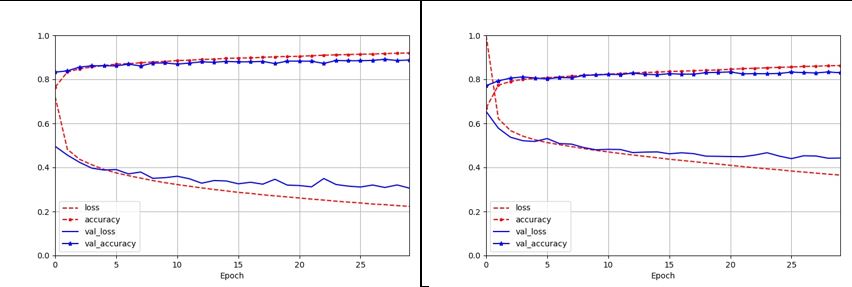} 
	\end{center}
	\caption{Comparison of learning curves with \cite[Chap. 10]{geron} algorithms. Plaintext dataset (left), HbHAI-protected version (right)} 
\end{figure}
We first use the same classic algorithm of TensorFlow/Keras (ANN/Multilayer Perceptron, three layers) proposed for MNIST-Zalando dataset by A. Géron \cite[Chap. 10]{geron}, without any modification and with the same parameters. 

We obtain a validation accuracy of $0.8188$ versus $0.8854$ for dataset without HbHAI. Figure~\ref{fig1} shows the two learning curves produced. So the results fall short of expectations.

%---------------------------------------------------------------------------------------------------------
\subsubsection{Analysis ($\delta = 3$) - 2-step Random Forest Approach}
\begin{comment}
\begin{figure}[h]\label{fig2}
	\begin{center} 
		\begin{tabular}{cc}
			Step 1                            &                  Step 2 \\
			\includegraphics[width=5.5cm]{confusionmatrixstep1.JPG}  & 	\includegraphics[width=5.5cm]{confusionmatrixstep2.JPG} 
		\end{tabular}
	\end{center}
	\caption{Confusion matrices. Step 1 on the left, step 2 on the right} 
\end{figure}
\end{comment}
In order to understand why the previous result were not as good as expected, we then perform a deep analysis to understand HbHAI data (which is a classical approach in data analysis/AI).

Whatever the method we used, we have observed that class 6 of the dataset was generally badly predicted. So we set up a 2-step detection algorithm based on decision trees \cite[Chap. 6]{geron} and Random Forests \cite[Chap. 7]{geron} from scikit.learn. We have proceeded as follows:
\begin{itemize}
	\item \textbf{Step 1}.- We used a Random Forest Model (1,500 decision trees). Class 6 is excluded. We obtained a validation accuracy of 90 \%.
	\item \textbf{Step 2}.- Than we used a second Random Forest Model (1,500 decision trees) to specifically process class 6 by using of an empiric decision threshold below which data are decided in class 6. 
\end{itemize} 
As a result, we got a final validation accuracy of $0.9534$.
%-----------------------------------------------------------------------------
\subsubsection{Analysis ($\delta = 6$) - Overall Results}
We ten considered the second version of the dataset with a compression rate of $\delta = 6$. The results can be summarized as follows: 
\begin{itemize}
	\item TensorFlow/Keras (ANN/Multilayer Perceptron, three layers) with A. Géron \cite[Chap. 10]{geron} algorithm, without any modification and with the same parameters, gives very unsatisfying results (validation accuracy around $0.7$).
	\item The data analysis showed that classes 2, 4 and 6 are badly predicted (Random Forest with scikit.learn). We have a first model $\mathcal{M}_1$).
	\item A cecond model $\mathcal{M}_2$ has been set up to decide specifically classes 2, 4 and 6.
	\item Finally we apply $\mathcal{M}_1$ and each time class 2, 4 or 6 is decided, we apply $\mathcal{M}_2$ for error checking/validation.
\end{itemize}
As final result we obtain a validation accuracy of Final validation accuracy of $0.938$ 
%--------------------------------------------------------------------
\subsubsection{Results Comparison}
A lot of researchers have proposed methods and results on Zalando's Fashion-MNIST \cite{dataset2}. We therefore compared our results on the HbHAI version of the Zalando dataset with the detection results of 38 researchers on the original version of the dataset (in plaintext). 
For $\delta = 3$ we obtain on HbHAI-protected data the 6$^{th}$ best result among 39 results (including ours)
\[97.5 > 96.7 > 96.3 > 95.9 > 95.4 > \color{red}{95.34}\color{black} > \ldots\]
As for $\delta = 6$ we obtain the 19$^{th}$ best result among 39 results. 
As a conclusion for $\delta \leq 6$, HbHAI techniques do not prevent to obtain excellent AI results with off-the-shelf software or algorithms.

Regarding the computing time results, we can summarize as follows:		
\begin{itemize}
\item We did not re-implement classic algorithm from scratch.
\item The classic tools, software or libraries (scikit.learn, TensorFlow\ldots) introduce a big overhead.
\item Computing time reduction when processing HbHAI-protected datasets is of at least 18~\% in average.
\item Writing dedicated, optimized is likely to provide far larger reduction, closer to the theory. 
\end{itemize} 

%%%%%%%%%%%%%%%%%%%%%%%%%%%%%%%%%%%%%%%%%%%%%%%%%%%%%%%%%%
\section{Conclusion \& Future Works} \label{sec4} 
In this paper we have presented the first evaluation of a new, disruptive-claimed approach of homomorphic encryption for AI. We expect that other researchers will perform their own evaluation once the datasets are publicly available.

Are HbHAI techniques performances and security claims confirmed? Our main results are the following ones:
\begin{itemize}
	\item In terms of security, only a black-box analysis was possible (no access to the core algorithm and the core primitives). Having each individual of datasets under both plaintext and encrypted forms, we however did not find any way to associate HbHAI-protected data with corresponding plaintext data. The effect of secret-based compression seems to be very secure. Only a white-box analysis (knowledge of the HbHAI technique itself) can allow the rigorous security assessment.
	\item We observed that HbHAI indeed provides an efficient and real preservation of AI algorithms efficiency/accuracy on HbHAI-protected data. We confirm that no information loss has occurred and that models are very well preserved.
	\item HbHAI technique really enables to work with "off-the-shelf" software/tools/libraries without modification (up to parameters). But the clear potential and power of HbHAI can really be exploited with a dedicated, optimized implementation of classic AI algorithms (as we did for the first dataset).
	\item In terms of computing performances and data size reduction, we confirm that the speed-up is compliant with data size reduction with dedicated, optimized classical AI algorithms. But with tools like scikit.learn or TensorFlow (the large overhead is due to some sort of internal inertia) only 20~\% computing time reduction has been measured. However is remains a huge speed-up compared to classic FHE versions of AI algorithms (one-million times slower compared to processing time on plaintext data \cite{jkun}).
\end{itemize}  
\textbf{Based on the analysis of only two datasets}, we can say that globally HbHAI seems a very promising, indeed disruptive approach. We hope that more technical will be shared publicly in order to study it further. 
%\appendix

\begin{credits}
\subsubsection{\ackname}  We would like to thank Dr Jaagup Sepp for the trust he placed in us by sharing the datasets before publication as well as the initial technical data on HbHAI. We would also like to thank him for allowing me to re-use part of his forthcoming article~\cite{sepp} so that this article is self-contained.

\subsubsection{\discintname}
There are no ethical issues. The author does not have any competing interest of any kind. This research work was entirely self-financed. HbHAI techniques are the exclusive property of Hope4Sec.
\end{credits}


\begin{thebibliography}{50}
\bibitem{he4ds} Allon Adir, Ehud Aharoni, Nir Drucker, Ronen Levy, Hayim Shaul and Omri Soceanu (2024). \textit{Homomorphic Encryption for Data Science (HE4DS)}. Springer Nature Switzerland AG.
	
\bibitem{comp4} Fono Louis Aim\'{e}, Yancho Basil Wiryen, Noumsi Auguste Vigny and Mvogo Joseph  Ngono (2024). Leveraging TenSEAL: A Comparative Study of BFV and CKKS Schemes for Training ML Models on Encrypted IoT Data. Int. J. Inf. Sec. Priv., 18(1), pp. 1--17, \url{https://doi.org/10.4018/IJISP.356402}.

\bibitem{tenseal} Ayoub Benaissa, Bilal Retiat, Bogdan Cebere and Alaa Eddine Belfedhal (2021). \textit{Tenseal: A library for encrypted tensor operations using homomorphic encryption}. ArXiv preprint arXiv:2104.03152. \url{https://arxiv.org/abs/2104.03152} Last accessed January 24$^{th}$.
	
\bibitem{ccks} Jung Hee Cheon, Andrey Kim, Miran Kim, and Yong Soo Song (2017). Homomorphic Encryption for Arithmetic of Approximate Numbers. In: 23$^{rd}$rd International Conference on the Theory and Applications of Cryptology and Information Security (AsiaCrypt), Tsuyoshi Takagi and Thomas Peyrin (Eds.). Lecture Notes in Computer Science 10625, Springer Verlag, pp. 409--437.

\bibitem{tfhelib} Ilaria Chillotti, Nicolas Gama, Mariya Georgieva and Malika Izabach{\`e}ne (2016). {TFHE}: Fast Fully Homomorphic Encryption Library. Available on \url{https://tfhe.github.io/tfhe/}.

\bibitem{fheinjatt} Ilaria Chillotti, Nicolas Gama and Louis Goubin (2016). Attacking {FHE}-based applications by software fault injections. Cryptology ePrint Archive, Paper 2016/1164, \url{https://eprint.iacr.org/2016/1164}. Last accessed April 16$^{th}$, 2025.

\bibitem{tfhe}  Ilaria Chillotti, Nicolas Gama, Mariya Georgieva and Malika Izabach{\`e}ne (2020). TFHE: Fast Fully Homomorphic Encryption over the Torus. Journal of Cryptology, volume 33, pages 34–91. Also available on \url{https://eprint.iacr.org/2018/421}.

\bibitem{comp2} Thi Van Thao Doan, Mohamed-Lamine Messai, Gérald Gavin and Jérôme Darmont (2022). A Survey on Implementations of Homomorphic
Encryption Schemes. PREPRINT (Version 2) available at Research Square \url{https://doi.org/10.21203/rs.3.rs-2018739/v2}. Last accessed January 24$^{th}$, 2025. 

\bibitem{dolev} Danny Dolev, Cynthia Dwork, Moni Naor (2000). Nonmalleable Cryptography. SIAM Journal on Computing. \textbf{30 (2)}, 391-–437.

\bibitem{comp5} Goran Đorđević, Milan Marković and Pavle Vuletić (2021). Performance comparison of homomorphic encryption scheme implementations. In: Proceedings of the IcETRAN 2021, pp. 514--520, available on \url{https://www.etran.rs/2021/zbornik/Papers/104_RTI_2.5.pdf}, last accessed December 12$^{th}$.

\bibitem{bfv} Junfeng Fan and Frederik Vercauteren (2012). Somewhat Practical Fully Homomorphic Encryption. Cryptology ePrint Archive, Paper 2012/144. \url{https://eprint.iacr.org/2012/144 https://eprint.iacr.org/2012/144}. Last accessed January 20$^{th}$, 2025.

\bibitem{filiol0} Eric Filiol (2025). New Proposal for Homomorphic AI. International Conference on the AI Revolution: Research, Ethics, and Society 
(AIR-RES 2025), April 14$^{}th$-16$^{th}$, 2025, Las Vegas, proceedings to be published by Springer. 

\bibitem{filiol1} Eric Filiol (2025). Technical Evaluation Hash-based Homomorphic AI. To be presented at CyberWiseCon 2025 \url{https://cyberwisecon.eu/}, May 20$^{th}$-23$^{rd}$, 2025, Vilnius, Lithuania. 

\bibitem{fix1951} Evelyn Fix and Joseph L. Hodges. Discriminatory Analysis, Non-parametric Discrimination: Consistency Properties. Technical Report 4, USAF School of Aviation Medicine, Randolph Field, 1951.

\bibitem{fhe} Craig Gentry (2009). A fully homomorphic encryption scheme. In: Proceedings of the 41$^{st}$ Annual ACM Symposium on Theory of Computing. ACM, pp. 169--178.

\bibitem{geron} Aurélien Géron. Hands-On Machine Learning with Scikit-Learn, Keras \& TensorFlow. Third ed., O'Reilly, 2022.

\bibitem{gmp} Torbjörn Granlund. The GNU Multiple Precision Arithmetic Library. Edition 6.3.0, \url{gmplib.org}, 2023.

\bibitem{helib} Homenc (2019). \textit{HElib: An open-source library implementing homomorphic encryption}. \url{https://github.com/homenc/HElib}. Last accessed January 24$^{th}$. 

\bibitem{openfhe} HPDIC Lab (2023). \textit{OpenFHE - Open-Source Fully Homomorphic Encryption Library v1.0.3}. \url{https://github.com/hpdic/openfhe-development}. Last accessed January 24$^{th}$.

\bibitem{huang1998} Zhexue Huang. Extensions to the k-Means Algorithm for Clustering Large Data Sets with Categorical Values. Data Mining and Knowledge Discovery 2, 283–304 (1998). https://doi.org/10.1023/A:1009769707641.

\bibitem{iea} International Energy Agency: World Energy Outlook 2024 (2024). \url{https://www.iea.org/reports/world-energy-outlook-2024}, last accessed 2025/01/02.

\bibitem{ieee1} IEEE Digital Privacy: What Is Homomorphic Encryption? (2024). \url{https://digitalprivacy.ieee.org/publications/topics/what-is-homomorphic-encryption}, last accessed 2025/01/02.

\bibitem{comp1} Ilia Iliashenko and Vincent Zucca (2021). Faster homomorphic comparison operations for BGV and BFV.
	author={}, Proceedings on Privacy Enhancing Technologies, volume 2021, pp. 246--264.

\bibitem{kostrikin} Aleksei Ivanovich Kostrikin (1982). Introduction to Algebra. Universitext Serie, Springer Verlag, New York, Heidelberg, Berlin.

\bibitem{comptfhe} Lei Jiang and Lei Ju (2022). FHEBench: Benchmarking Fully Homomorphic Encryption Schemes. ArXiv preprint 2203.00728, \url{https://arxiv.org/abs/2203.00728}, last accessed April 15$^{th}$, 2025. 

\bibitem{jkun} Jeremy Kun (2024). A High-Level Technical Overview of Fully Homomorphic Encryption. \url{https://www.jeremykun.com/2024/05/04/fhe-overview/}, last accessed December 14$^{th}$, 2024.
	
\bibitem{mnist} Yann LeCun, Corinna Cortes and Chris Burges (1998). The MNIST Database of Handwritten Digits. \url{https://yann.lecun.com/exdb/mnist/}, last accessed 2024/11/25.

\bibitem{leinster} Tom Leinster (2016). Basic Category Theory. Cambridge Studies in Advanced Mathematics, Cambridge University Press. A preprint version is available on \url{https://arxiv.org/abs/1612.09375}.

\bibitem{lin} Wilfred W. K. Lin (2023). Challenges of Homomorphic Encryption. \url{https://www.researchgate.net/publication/370050235_Challenges_of_Homomorphic_encryption}, last accessed 2024/11/23.

\bibitem{lobo2020} Matheus P. Lobo. Homomorphisms: A Concise Approach. OSF Preprints, \url{osf.io/7pgcs_v1}, 2020.

\bibitem{rlwe} Vadim Lyubashevsky, Chris Peikert, and Oded Regev (2010). On ideal lattices and learning with errors over rings. In: Advances in Cryptology – EUROCRYPT 2010. Lecture Notes in Computer Science 6110, Springer, Berlin, Heidelberg, pp. 1–23.

\bibitem{mccullagh} Peter McCullagh (2002). What Is a Statistical Model? The Annals of Statistics, Vol. 30, No. 5, pp. 1225-1267.

\bibitem{hac} Alfred J. Menezes, Paul C. van Oorschot and Scott A. Vanstone (2007). \textit{Handbook of Applied Cryptography}. CRC Press. Available at \url{http://www.cacr.math.uwaterloo.ca/hac/}

\bibitem{seal} Microsoft (2018). \textit{Microsoft SEAL (Simple Encrypted Arithmetic Library)}. \url{https://github.com/microsoft/SEAL}. Last accessed January 24$^{th}$, 2025. 

\bibitem{philmun} Paul Muncaster (2024). US Data Breach Victim Numbers Surge 1170~\% Annually. Info Security Magazine (2024).  \url{https://www.infosecurity-magazine.com/news/us-data-breach-victims-surge-1170/}, last accessed 2024/12/19.

\bibitem{rand1971} William M. Rand. Objective Criteria for the Evaluation of Clustering Methods. Journal of the American Statistical Association
Vol. 66, No. 336, pp. 846-850, 1971. 

\bibitem{schneider} Jonas Schneider (2024). Awesome Homomorphic Encryption. \url{https://github.com/jonaschn/awesome-he}, last accessed 2025/01/02.

\bibitem{sepp} Jaagup Sepp. Datasets for Hash-based Homomorphic AI with Variable Compression Rate. World Congress in Computer Science, Computer Engineering, \& Applied Computing (CSCE'25) - The 27th International Conference on Artificial Intelligence (ICAI'25), July 21$^{st}$-24$^{th}$, 2025, Las Vegas, USA. Proceedings to be published by Springer.   

\bibitem{comp3} Arisa Tsuji and Masato Oguchi (2024). Comparison of FHE Schemes and Libraries for Efficient Cryptographic Processing. In: Proceedings of the 2024 International Conference on Computing, Networking and Communications (ICNC): Edge Computing, Cloud Computing and Big Data, pp. 584--590, \url{http://www.conf-icnc.org/2024/papers/p584-tsuji.pdf?title=PDF+file}. Last accessed January 24$^{th}$, 2025.

\bibitem{zama} Michael Walter (2022). On Side-Channel and {CVO} Attacks against {TFHE} and {FHEW}. Cryptology {ePrint} Archive, Paper 2022/1722,
\url{https://eprint.iacr.org/2022/1722}. Last accessed April 15$^{th}$, 2025.
 
\bibitem{dataset2} Zalando Research: Fashion-MNIST (2017). \url{https://github.com/zalandoresearch/fashion-mnist}, last accessed 2024/11/05.

\bibitem{hades} Dongfang Zhao (2024). Hades: Homomorphic Augmented Decryption for Efficient Symbol-comparison -- A Database's Perspective. ArXiv preprint 2412.19980, \url{https://arxiv.org/abs/2412.19980}, last accessed January 24$^{th}$, 2025.
	
\end{thebibliography}
\end{document}